\documentclass[twocolumn,showpacs,preprintnumbers,amsmath,amssymb,prb,superscriptaddress]{revtex4}

\usepackage{graphicx}
\usepackage{color}
\usepackage{pstricks}
\usepackage{epsfig}
\usepackage{overpic}

\begin{document}

\title{
Nonequilibrium transport properties of a double quantum dot in the Kondo regime}
\author{D.~Breyel$^1$ and A.~Komnik}
\affiliation{Institut f\"ur Theoretische Physik, Universit\"at Heidelberg,\\
 Philosophenweg 19, D-69120 Heidelberg, Germany}
\date{\today}

\begin{abstract}
We analyze the nonequilibrium transport properties of a parallel double quantum dot in terms of its full counting statistics (FCS). The parameters of the setup are
assumed to be such that both subsystems are driven into the Kondo regime. After a series of transformations the Hamiltonian is then mapped onto a Majorana resonant
level model, which effectively describes the Toulouse point of the respective double impurity two-terminal Kondo model. Its FCS is then obtained at arbitrary
constellation of voltage, temperature, and local magnetic fields. We identify two different transport processes corresponding to single electron tunneling as well
as an electron pair process and give the respective effective transport coefficients. In the most universal linear response regime the FCS turns out to be of a
binomial shape with an effective transmission coefficient. Furthermore, we find a complete transport suppression (antiresonance) at a certain parameter constellation,
which is similar to the one found in the noninteracting quantum dots. By an explicit expansion around the Toulouse point we show that the antiresonance is universal and
should be observable in the generic Kondo dot setup.
We discuss experimental implications of our predictions as well as possible routes for generalizations of our approach.

\end{abstract}


\maketitle

\section{Introduction}

Quantum dot (QD) is an important paradigm in many research fields. One of the most important application areas of this concept is the quantum transport. Here the QD
often represents a basic building element of micro- and nanoelectronic circuitry\cite{Nazarov2009}. For the applications it is not only important to understand the
physics of individual devices but also of more complicated arrangements of QDs. The focus of our study is the structure with the next to single QD complexity level --
which is besides slight modifications and similar geometries\cite{sela:125110, PhysRevB.51.8287, 0295-5075-92-1-10002, PhysRevLett.99.036807} a double QD structure
in parallel geometry.\cite{RevModPhys.75.1, 0953-8984-19-25-256205, refId, DiasdaSilva20081002, PhysRevB.77.045309, PhysRevB.78.153304, PhysRevB.81.113402, PhysRevLett.97.247207, PhysRevLett.106.147202,Sela2009}


The most widespread modeling strategy for a QD is the bottom-up approach. The structure is essentially assumed to be a zero-dimensional object, which is modeled by a
single spin-degenerate fermionic level. This ``resonant level'' model is then coupled via particle exchange to source and drain electrodes as well as electostatically
to a gate electrode which tunes the resonant level energy thereby controlling the transmission coefficient of the structure. However, because of a confined geometry
one has to include electron interaction terms at least in the minimal form by introduction of finite energy cost for the double population of the QD. This leads
directly to the celebrated Anderson impurity model (AIM)\cite{Anderson1961,Hewson1997} which was already applied to parallel double dot structures.
\cite{PhysRevLett.99.076602, PhysRevLett.96.146801, PhysRevB.83.115310} It shows up an enormous manifold of very different transport regimes, for many of which no exact analytical
results in nonequilibrium are available.

Probably one of the most interesting cases is the Kondo regime, when the QD is populated by a single electron and energetic cost of double population as well as
of emptying the dot are so high that the transport channel involving only single-electron tunneling events is blocked.\cite{GlazmanAnkara} Then only virtual double
population is allowed and the remaining transport channel is the spin-flip tunneling. It has very profound consequences on the transport characteristics of the systems.
For instance, the nonlinear $I(V)$ as well as the shot noise turns out to contain two different transport channels -- single electron tunneling and a pair
process,\cite{Meir2002,Sela2006,Golub2006} which can be seen explicitly in their FCS.\cite{Gogolin2006,AndersonFCS} The resulting highly nontrivial Fano
factor $5e/3$ has already been verified experimentally.\cite{Zarchin2008} Due to the additional degrees of freedom of double QDs it might, at least in principle,
be possible for the electrons to be transferred in groups of not only two but also of three and even four particles. FCS is most useful in answering such questions.

Although the low-energy transport characteristics of the Kondo regime of double QDs are by now fairly good understood, the nonlinear response and especially the
FCS of such systems are notoriously difficult to obtain. There is, however, a parameter constellation, which is on the one hand a non-trivial one in a sense that
it captures most of the relevant physics of the Kondo limit, and which on the other hand allows for a complete analytical solution with elementary means. This is
known under the name of Toulouse point solutions\cite{Toulouse1969a,Emery1992,Schiller1998,Gogolin1998}.

A similar calculation for the double QD setup, being an analytic solution, would yield invaluable information about the transport properties of the system as
well as become an important benchmark for other approaches. To the best of our knowledge this kind of calculation was not yet attempted. There are several
obstacles which need to be circumvented. The first issue is the validity of the Hamiltonian in which two different localized spins are simultaneously \emph{locally}
coupled to two electrodes. This part of the program was successfully mastered in Ref.~[\onlinecite{Zitko2006}]. The second question is that of the principal
applicability of the sequence of transformations which are usually used in order to obtain an essentially quadratic Hamiltonian. As is argued in the next section
one can indeed apply them after a minor adjustment. The last issue is slightly more technical and concerns the method for FCS calculation, how one has to introduce
the counting fields in order to obtain meaningful results. This is discussed with all necessary details in Section \ref{modelandcalculationmethod}. In Section
\ref{fcsofdd} we present the results and
discuss the emerging physics. Conclusions Section summarizes our findings and gives directions for future research.

\section{Model and calculation method}
\label{modelandcalculationmethod}

\begin{center}
 \begin{figure}
 \includegraphics[width=\columnwidth,bb=14 14 771 332,keepaspectratio=true]{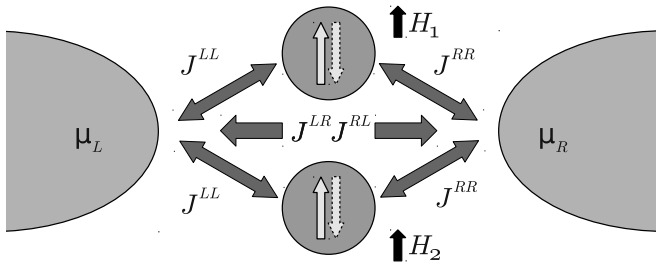}
 \caption{Parallel double quantum dot. The two quantum dots are coupled symmetrically (under exchange of the dots) to metallic leads. $H_{1,2}$ (small arrows) are the in-plane magnetic fields. The dot spins are represented
 by larger arrows. The coupling is represented by the doubly headed arrows.}
 \label{gra:system}
 \end{figure}
\end{center}

The simplest meaningful starting point is the double impurity Anderson model with two electrodes.
It is shown in Ref.~[\onlinecite{Zitko2006}] that by a dedicated Schrieffer-Wolff transformation\cite{Schrieffer1966} the low energy
sector is that of two magnetic impurities coupled by exchange terms to the spin densities in the electrodes (as long as only one
electron populates the dot). This situation is shown in Fig.~\ref{gra:system} which also serves to explain the notation conventions.
We want to emphasize at this point that the applied magnetic fields, which are independent on each of the dots, are aligned parallel to
the plane of the structure and therefore do not create an Aharonov-Bohm phase. The applied bias voltage $V$ is indicated by the different
chemical potentials on the leads: $ \mu_L - \mu_R = eV. $ Then the corresponding Hamiltonian is given by

\begin{equation}
 \mathcal{H} = \mathcal{H}_{\text{kin}} + \mathcal{H}_{\text{int}} + \mathcal{H}_{\text{mag}},
 \label{eqn:hamiltonian}
\end{equation}

\noindent where the three parts are given by (we use units in which $e=\hbar = v_F = 1$, where $v_F$ is the Fermi velocity in the electrodes)

\begin{gather}
 \mathcal{H}_{\text{kin}} = i \sum\limits_{\alpha=L,R} \sum\limits_{\sigma = \uparrow,\downarrow} \int \psi_{\alpha \sigma}^\dagger(x) \partial_x \psi_{\alpha \sigma}(x) \text{d}x \, , \\
 \mathcal{H}_{\text{int}} = \sum\limits_{\alpha,\beta=L,R} \sum\limits_{\lambda = x,y,z} J^{\alpha \beta}_\lambda s^\lambda_{\alpha \beta}(x=0) \left( \tau_1^\lambda + \tau_2^\lambda \right) \, , \\
 \mathcal{H}_{\text{mag}} = -\mu_\text{B} g_{\text{imp}}\left( H_1 \tau_1^z + H_2 \tau_2^z \right) = -\sum\limits_{j=1,2}\Delta_j \tau_j^z.
\end{gather}
$\psi^\dagger_{\alpha \delta}(x)$ is the fermionic field creation operators in the electrode $\alpha$. The bare (without impurity coupling) electron dispersion in the leads is linearized around the Fermi edge as we are only interested in the low energy sector of the system.
The interaction term includes the generalized spin densities $s^\lambda_{\alpha \beta}(x) = \frac{1}{2} \psi^\dagger_{\alpha \delta}(x) (\sigma)^\lambda_{\delta \delta^\prime} \psi_{\beta \delta^{\prime}}(x)$. $\mu_B$ is the Bohr's magneton and $g_{\rm imp}$ is the impurity Land\'e factor.
We restrict our calculations to the cases where the coupling constants fulfill the restrictions
\begin{eqnarray}               \label{restrictions1}
 J^{\alpha \beta}_x = J^{\alpha \beta}_y = J^{\alpha \beta}_\perp, \; J^{LR}_z = J^{RL}_z =0 \notag \\
 \text{and} \; J^{LL}_z = J^{RR}_z = J_z
\end{eqnarray}

\noindent first discussed in [\onlinecite{Schiller1998}]. We employ the strategy outlined there and proceed by performing a number of transformations in order to map the Hamiltonian \eqref{eqn:hamiltonian} onto the one of a noninteracting system.
First the Hamiltonian is bosonized using the bosonization identity\cite{Gogolin1998,Giamarchi2004}
\begin{equation}
 \psi_{\alpha \sigma} (x) = \frac{e^{i \varphi_{\alpha \sigma}}}{\sqrt{2 \pi a}}e^{- i \phi_{\alpha \sigma}}
 \label{eqn:bosonizationidentity} \, ,
\end{equation}
where $a$ is the lattice constant of the underlying lattice model.
 In the next step we introduce the bosonic fields $\phi_c,\phi_s,\phi_f$ and $\phi_{sf}$ (corresponding to ``charge'', ``spin'', ``flavour'' and ``spin-flavour'' channels)
as linear combinations of the bosonic fields appearing in \eqref{eqn:bosonizationidentity}:
\begin{equation}
 \phi_{c,s,f,sf}= \frac{1}{2} (\phi_{L \uparrow}  {_{\pm}^{\pm}}  \phi_{L \downarrow} {_{-}^{+}}  \phi_{R \uparrow} {_{\mp}^{\pm}} \phi_{R \downarrow}).
\end{equation}
 We apply a unitary transformation which is known to change the scaling dimensions of the coupling thus enabling its refermionization.\cite{Gogolin1998} In the spirit of Ref.~[\onlinecite{Kawaguchi2009}] we construct it as a product of of two
Emery--Kivelson rotations\cite{Emery1992}
\begin{equation}
 U= e^{\frac{i}{2} \chi_s (\tau_1^z + \tau_2^z)}.
\end{equation}

\noindent around the $z$-axes of the spins. Here $\chi_s$ denotes the combination of the bosonic field with its corresponding Klein factor field $\chi_\nu =\phi_\nu (0) - \varphi_\nu$. 
One can define new fermionic fields
\begin{align}
 & \psi_{f} = \frac{e^{i \pi (d^\dagger_1 + d^\dagger_2 )(d_1 + d_2)}}{\sqrt{2 \pi a}} e^{-i ( \phi_f - \varphi_f)},\\ & \psi_{sf} = \frac{e^{i \pi (d^\dagger_1 + d^\dagger_2 )(d_1 + d_2)}}{\sqrt{2 \pi a}} e^{-i ( \phi_{sf} - \varphi_{sf})} \, ,
\end{align}
\noindent which allow for a refermionization of the Hamiltonian yielding

\begin{multline}
 \mathcal{H}^\prime = \mathcal{H}_{\text{kin}} + \left( 2 \sqrt{2 \pi a} \right)^{(-1)} \\
  \times \Big[ J^+ (\psi^\dagger_{sf} + \psi_{sf})(d_1^{\dagger} - d_1 + d_2^{\dagger} - d_2 ) \\
  + J^{LR} (\psi^\dagger_{f} - \psi_{f})(d_1^{\dagger} + d_1 + d_2^{\dagger} + d_2 ) \\
  + J^{-} (\psi^\dagger_{sf} - \psi_{sf})(d_1^{\dagger} + d_1 + d_2^{\dagger} + d_2 ) \Big] \\
  + \sum\limits_{j=1,2} \left[ \mu_B g H_j - (J_z - 2\pi) :\psi_s^\dagger(0) \psi_s(0): \right] (d_j^\dagger d_j - 1/2) ,
\end{multline}

\noindent where the free fermion field $\psi_s(x)$ has to be evaluated at $x=0$. Now we simplify this expression by introducing Majorana fermions for the dot and lead fermions
which are defined in the following way:

\begin{align*}
 a_j & = \frac{1}{\sqrt{2}} (d_j^\dagger + d_j), & \eta_\nu(x) & = \frac{1}{\sqrt{2}} [\psi_\nu^\dagger(x) + \psi_\nu(x)], \\
 b_j & = \frac{i}{\sqrt{2}} (d_j - d_j^\dagger), & \xi_\nu(x) & = \frac{i}{\sqrt{2}}[\psi_\nu(x) - \psi_\nu^\dagger (x)] \, .
\end{align*}

\noindent For purposes that will become clear later in this work we want to define a new parameter $\gamma$ to be $\gamma := J_z - 2 \pi$. In this new
notation the transformed Hamiltonian reads

\begin{eqnarray}
 \mathcal{H}^\prime &=& \mathcal{H}_\text{kin} -i \Delta_1 a_1b_1 -i \Delta_2 a_2 b_2 + i K_{LR} \xi_f(a_1 + a_2) \nonumber \\ &+& i K_+ \eta_{sf}(b_1 + b_2) + i K_- \xi_{sf} (a_1 + a_2)
 \nonumber \\
 &+& \gamma :\psi_s^\dagger(0) \psi_s(0): (\tau_1^z + \tau_2^z).
 \label{eqn:transformedhamilton}
\end{eqnarray}

\noindent We note that after the transformations the coupling constants $K_j$ as well as the fermionic fields $\psi_\nu$ have the physical dimension $\sqrt{\text{energy}}$.
This can be seen in the definitions

\begin{align}
 & J^{\pm} = \frac{1}{2} ( J^{LL}_\perp \pm J^{RR}_\perp ), & & K_j = \frac{J^j}{\sqrt{2 \pi a}},
\end{align}
\noindent where $a$ is the constant that corresponds to the lattice spacing that already appeared in
Eq.~\eqref{eqn:bosonizationidentity}.

In its new form the Hamiltonian, which does not contain any approximations or simplifications yet, possesses only one term which is not quadratic in the fermionic fields.
In fact, in the case $\gamma = 0$ or $J_z = 2\pi$ it becomes purely quadratic and thus exactly solvable by elementary means. This particular point in the parameter space is referred
to as the \emph{Toulouse point}.\cite{Gogolin1998} Despite its relative simplicity, Toulouse point solution carries all features of the generic Kondo effect because
$J_z = 2\pi$ corresponds to rather \emph{strong} correlations. The conventional strategy is to solve the $\gamma=0$ case first and then analyze the robustness of the
solution beyond that point by an expansion around the Toulouse point.\cite{Majumdar1998,AndersonFCS,Schiller1998} We follow this path in the remainder of the paper. \\

At this point we would like to get back to our earlier restriction of the magnetic fields being aligned to the plane of the structure
to ensure that no Aharonov-Bohm phase (AB phase) is generated. Now, having performed the bosonization, rotation and refermionization procedure, we
may ask if it is possible to include this phase in this formalism. The idea then is to break the symmetry $J^{LR} = J^{RL}$ and to equip
each of the corresponding terms in the Hamiltonian with an appropriate phase $e^{\pm i \alpha/2}$. The other terms are not affected
by the AB phase. Executing the same steps up to the rotation leads to an expression that cannot be refermionized in the same manner as
above because of the Klein factors. They lead to extra factors of $\tau_{1,2}^z$ in the refermionized expression and therefore the
Hamiltonian cannot be mapped onto a noninteracting system. That is why we restrict ourselves to the in-plane magnetic fields.

\section{FCS of the double QD system}
\label{fcsofdd}

One of the most generic transport properties of QDs is the FCS. It is usually obtained in form of the cumulant generating function (CGF) $\ln \chi(\lambda)$, which,
being derived $n$ times with respect to the counting field $\lambda$ yields the $n$th cumulant (irreducible moment) of the probability to measure the transmission of
charge $Q$ during a very long measurement time ${\cal T}$,\cite{lll,Nazarov2009}

%
%

\begin{equation}
 \langle \! \langle Q^n \rangle \! \rangle = \frac{1}{i^n} \frac{\partial^n}{\partial \lambda^n} \ln  \chi(\lambda) \, .
 \label{eq:cumulants}
\end{equation}

\noindent The current through the constriction is then found from $I(V) = \langle \! \langle Q \rangle \! \rangle/{\cal T} $, the shot
noise is related to the second cumulant at zero temperature $S(V) = \langle \! \langle Q^2 \rangle \! \rangle/{\cal T}$ etc. It is also useful to construct a Fano factor, which is given by the (in our convention dimensionless) ratio of two lowest order cumulants,\cite{PhysRev.72.26,Nazarov2009}
\begin{equation}
 F(V) = \frac{S(V)}{2 I(V)} \, .
 \label{eqn:fanofactor}
\end{equation} It is related to the eminent Schottky formula, which allows (at least in principle) a measurement of the charge of current carrying excitations, see e.~g. Ref.~[\onlinecite{Zarchin2008}]. The main advantage of the CGF
is therefore that all cumulants are accessible via simple derivation with respect to $\lambda$ and setting $\lambda=0$ afterwards.  \\ 

There are several methods of CGF calculation. We shall follow the one presented in Ref.~[\onlinecite{Komnik2005}]. First the counting field $\lambda$ is introduced on the Keldysh contour as
\begin{equation}
 \lambda(t) = \theta(t)\theta(\mathcal{T}-t) \begin{cases} \lambda = \lambda_- \, , &  t \; \text{on forward path} \\ -\lambda = \lambda_+ \, , & t \; \text{on backward path} \end{cases}
\end{equation}
The charge counting operator
\begin{equation}
 T_\lambda = T_R e^{i \lambda(t)/2} + T_L e^{-i \lambda(t)/2}
\end{equation}
is constructed from $T_{R(L)}$ parts of the Hamiltonian, which are responsible for particle transport from the right electrode to the left one and vice versa. In the present situation it amounts to the decoration of tunneling terms $\psi^\dagger_{L\sigma}\psi_{R\sigma^{\prime}}$ ($\psi^\dagger_{R\sigma}\psi_{L\sigma^{\prime}}$) in
the starting Hamiltonian \eqref{eqn:hamiltonian} by factors
$ e^{-i \lambda(t)/2} $ ($ e^{i \lambda(t)/2} $), respectively.

The CGF can then be derived using the formula\cite{levitov04,Levitov1993}
\begin{equation}
 \chi(\lambda) = \langle T_{C} e^{-i \int_{C} T_\lambda(t) \text{d} t} \rangle \, ,
\end{equation}
where $T_{C}$ is the Keldysh contour ordering operator and the average is taken with respect to the full $\mathcal{H}$.
Although this average can be calculated directly, the \emph{adiabatic potential} method results in a much more compact algebra.\cite{Komnik2005} It turns out that due to the Feynman--Hellmann theorem\cite{Feynman1939} in the limit ${\cal T} \rightarrow \infty$ there is an identity
$\ln \chi(\lambda)  = -i \mathcal{T} \mathcal{U}(\lambda_-,\lambda_+)$, where the adiabatic potential is defined as
\begin{equation}
 \frac{\partial}{\partial \lambda_-} \mathcal{U}(\lambda_-,\lambda_+) = \left\langle \frac{\partial T_\lambda(t)}{\partial \lambda_-} \right\rangle_\lambda \, .
\end{equation}
In the Majorana representation the charge counting term is then given by
\begin{equation}
 T_\lambda (t) = i K_{LR} [\sin(\lambda/2) \, a_+ \eta_f + \cos(\lambda/2) \, a_+ \xi_f] \, ,
\end{equation}
where we introduced the new fields $a_\pm = (a_1 \pm a_2)/\sqrt{2}$. Because of the normalization of these new fields a factor
of $\sqrt{2}$ has to be absorbed in every coupling constant. Analogously to the procedure shown in Ref.~[\onlinecite{schmidt:235105}] we express the adiabatic potential in terms of Green's functions (GFs)
\begin{multline}
 \mathcal{U}(\lambda) = -i K_{LR}^2 \int \frac{\text{d}\omega}{2 \pi} \int \text{d} \lambda \left[\
D^{--}_{a_+ a_+}g^{--}_{\xi_f \eta_f} \right. \\ \left. + \sin(\lambda) D^{-+}_{a_+ a_+} g^{+-}_{\eta_f \eta_f} - \cos(\lambda) D^{-+}_{a_+ a_+} g^{+-}_{\xi_f \eta_f}\right] \, ,\
\end{multline}
where we used $\lambda = (\lambda_- - \lambda_+)/2$ and $D$ represent exact GFs whereas $g$ represent the GFs of zeroth order in coupling. Calculation of the exact GFs for the essentially quadratic Hamiltonian (we have set $\gamma=0$ in the whole of this section) is straightforward but rather lengthy. The final result for the CGF reads
\begin{align}
 \ln  \chi\left( \lambda \right)  & = \mathcal{T} \int\limits_{0}^{\infty} \frac{\text{d}\omega}{2 \pi} \ln\Big\{  1 + T_2(\omega) n_L(n_R-1) \left( e^{2 i \lambda} -1 \right) \notag \\
 & + T_1(\omega) \left[ n_R(n_F-1) + n_F(n_L-1) \right] \left( e^{-i \lambda} -1 \right) \notag \\
 & + T_1(\omega) \left[ n_L(n_F-1) + n_F(n_R-1) \right] \left( e^{i \lambda} -1 \right) \notag \\
 & + T_2(\omega) n_R(n_L-1) \left( e^{-2 i \lambda} -1 \right)  \Big\},
\label{eqn:CGF}
\end{align}
\noindent where $T_1(\omega)$ and $T_2(\omega)$ are effective transmission coefficients which are given by
\begin{equation}
 T_1(\omega) = \frac{\alpha_1}{\alpha_0} \; \; \text{and} \; \; T_2(\omega) = \frac{\alpha_2}{\alpha_0} \; \; \text{with}
\end{equation}
\begin{widetext}
 \begin{align*}
  \alpha_0 = & 32 K_+^2 [K_{LR}^2 + K_{-}^2] (\Delta_1 + \Delta_2)^2 (\omega^2 - \Delta_1 \Delta_2)^2 + 16 \omega^2 K_{+}^4 (\Delta_1^2 + \Delta_2^2 - 2 \omega^2)^2  \\
        & + [K_{LR}^2 + K_{-}^2]^2 \big\{ K_{+}^4 [(\Delta_1 + \Delta_2)^2 - 4 \omega^2]^2 + 16 \omega^2 (\Delta_1^2 + \Delta_2^2 - 2\omega^2)^2 \big\} + 256 (\Delta_1^2 - \omega^2)^2(\Delta_2^2 - \omega^2)^2, \\
  \alpha_1 = & 2 K_{LR}^2 \big\{ 16 K_{+}^2 (\Delta_1 + \Delta_2)^2 (\omega^2 - \Delta_1 \Delta_2)^2 + 16 K_{-}^2 \omega^2 (\Delta_1^2 + \Delta_2^2 - 2\omega^2)^2 + K_{+}^4 K_{-}^2 [(\Delta_1 + \Delta_2)^2 - 4 \omega^2]^2 \big\}, \\
  \alpha_2 = & K_{LR}^4 \big\{ K_{+}^4 [(\Delta_1 + \Delta_2)^2 - 4 \omega^2]^2 + 16 \omega^2 [\Delta_1^2 + \Delta_2^2 - 2 \omega^2]^2  \big\}.
 \end{align*}
\end{widetext}
In Eq.~\eqref{eqn:CGF} the factors $e^{\pm i \lambda}$ and $e^{\pm 2 i \lambda}$ distinguish between two different types of charge transfer processes: the first kind of processes transports
a single charge across the system while the second one transports two charges. The ``$\pm$'' indicates the direction of charge transport where ``+'' means ``in direction of applied voltage'' and ``--''
means ``against direction of applied voltage''. As expected, one can see that in the zero temperature limit the terms corresponding to the ``--'' vanish as the Fermi functions become Heaviside's step functions.
Unfortunately, at least in the chosen parameter range the FCS does not point towards collective processes involving three or four tunneling electrons. One reason for
that might be the absence of mutual many-particle correlations \emph{between} the dots.\cite{L'opez2005,Kashcheyevs2007a} An inclusion of such interactions leads to the Toulouse point break down so that
we have to postpone answering this question to a future work. \\
We also note that in the case of $\Delta_1 = 0 = \Delta_2$ our CGF, using the proper convention of notation, coincides with the result of [\onlinecite{AndersonFCS}] in the case of zero magnetic
field. This means that without any magnetic field applied the double Kondo impurity behaves just as a single one.

\subsection{Linear response regime $V,T \rightarrow 0$}
\label{linearresponseregime}

In this case we can expand our FCS in the
following way
\begin{equation}
 \ln \chi(\lambda) \approx \mathcal{T} \int\limits_0^V \frac{\text{d}\omega}{2 \pi} \, f(\omega)|_{\omega = 0} + \ldots ,
\end{equation}
\noindent where we write the integrand as a single function $f(\omega)$ for simplicity. Thus the integration over $\omega$ becomes trivial if
we limit ourselves to the zeroth expansion term. Then the FCS reads
\begin{multline}
 \ln \chi(\lambda)  = \frac{\mathcal{T} V}{2 \pi} \ln \big[ 1 + T_1(\omega=0)(e^{i \lambda} - 1)  \\ +   T_2(\omega=0) (e^{2 i \lambda} - 1)\big] \, ,
\end{multline}
\noindent which is equivalent to a manifestly binomial distribution\footnote{This result is consistent with the \emph{binomial} theorem put forward in Ref.~[\onlinecite{AndersonFCS}].}
\begin{align}
 \chi(\lambda) & = \ln \left[\ 1 + T_1(0)(e^{i \lambda} - 1) + T_2(0) (e^{2 i \lambda} - 1)\right]\ ^{\frac{\mathcal{T}V}{2\pi}} \notag \\
 & = \left[\ 1 + T_e (e^{i \lambda} - 1)  \right]\ ^{2 \frac{\mathcal{T}V}{2\pi}} \, .
\end{align}
$T_e = \sqrt{T_2(\omega=0)}$ is an effective transmission coefficient. In terms of the $\alpha_i$ introduced above the
transmission coefficients reduce to
\begin{align}
 \alpha_0 & = [ 16 \Delta_1^2 \Delta_2^2 + K_+^2 (K_+^2 + K_-^2) (\Delta_1 + \Delta_2) ]^2, \\
 \alpha_1 & = 2 (\Delta_1 + \Delta_2)^2 K_{LR}^2 K_+^2 [16 \Delta_1^2 \Delta_2^2 \notag \\
     & \hspace{3.3cm} + (\Delta_1 + \Delta_2)^2 K_-^2 K_+^2], \\
 \alpha_2 & = K_{LR}^4 K_+^4 (\Delta_1 + \Delta_2)^4.
\end{align}

At this point we would like to call the reader's attention to the fact that for the case $\Delta_1 = -\Delta_2$ both $T_1$ and $T_2$
vanish whatever the choice of coupling constants. This means that in this case the transport through the system is suppressed
and we observe an antiresonance at $\omega = 0$. This feature can be understood by the fact that for each
electron which spin-flip tunnels across the system its corresponding holes does the same via the other quantum dot. This effect is very similar to the one found in the
noninteracting double QDs.\cite{Hackenbroich1996,Kubala2003} (However, due to its different nature it does not allow to be used for spin filtering as suggested in
Ref.~[\onlinecite{Dahlhaus2010}]. See also Ref.~[\onlinecite{PhysRevB.66.125315}] for other interference effects.) Whether this feature pertains to the Toulouse point only is easily answered by a number of different perturbative expansions around this special point.\cite{Majumdar1998} We explicitly performed the lowest order perturbative expansion in $\gamma$.
This is a rather involved calculation, details of which we present in the Appendix. In the case of small bias voltages the transmission is dominated by the lowest order in $\omega$ contributions from the respective self-energy. It turns out, that the only constant term is generated by the Toulouse point terms $\gamma=0$, higher order $\gamma$-terms coming with higher powers of $\omega$. This can be understood as all Toulouse point correction terms being of \emph{inelastic} origin kicking in at finite energies. This is, of course, consistent with the scaling dimensions of the correction terms.

Very similar picture emerges upon loosening the restrictions (\ref{restrictions1}).\cite{Majumdar1998} That is why we expect the antiresonance to be robust and universal beyond the Toulouse point.
The only requirement for its realization is the fine-tuning of the magnetic fields to opposite values for both dots. We believe that this
is experimentally feasible by e.~g. applying inhomogeneous in-plane magnetic fields with finite spacial gradient in the setups which were used in Refs.~[\onlinecite{Holleitner2001,Wilhelm2002}].

\subsection{Finite voltage, zero temperature}

\begin{center}
 \begin{figure}
 \includegraphics[width=\columnwidth]{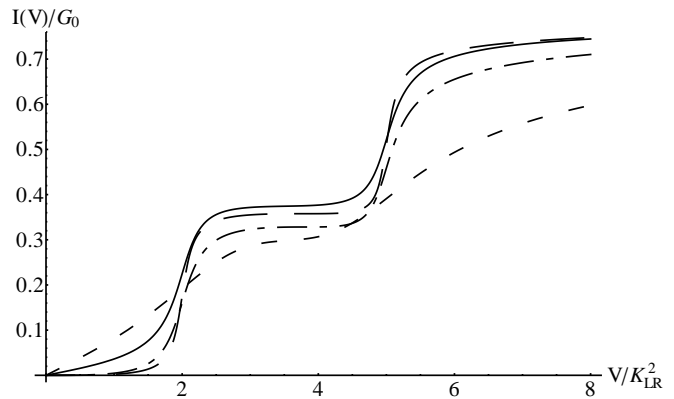} 
 \caption{The nonlinear $I(V)$ for the double QD setup at zero temperature. The plots show the current for $\Delta_1 = 2$ and $\Delta_2 = 5$. The different graphs correspond to $K_+ = 0$ and $ K_- = 1$
	  (dash-dotted line), $K_+ = 2 $ and $ K_- = 2$ (short dashed line),
	  $K_+ = 1$ and $K_- = 0$ (solid line) and $K_+ = 0$ and $K_- = 0$ (long dashed line).  All energies are measured in units of $K_{\rm LR}^2$. }
 \label{gra:current}
 \end{figure}
\end{center}

Eq.~\eqref{eqn:CGF} allows for the calculation of any cumulant desired for the given setup. At this point we would like to present some experimentally observable
quantities and discuss their specific features. In Fig.~\ref{gra:current} the electric current through the system is shown for four different sets of coupling constants while
the strength of the magnetic field is the same for all plots. The basic behaviour of all four curves is qualitatively identical: the current
 shows up two distinct plateaus as well as two steep shoulders before it saturates for a sufficiently large voltage. In any
case the two voltages where the current increases are equal to the strength of either of the magnetic fields. The reason for that kind of behavior is rather simple. Since the primary transport mechanism is the spin-flip tunneling it is suppressed for both dots at voltages smaller than the respective magnetic field. As soon as $V$ overcomes the smaller of the fields, the respective transport channel opens and there is a rapid increase of the current. The second step is then associated with the other field. The overall saturation of the current is related to the finite total spectral density of the constriction as required by the sum rules.

As far as the parameter dependence is concerned, see Fig.~\ref{gra:current}, the general trend that can be observed is that the shoulders of the
curves tend to smear out for larger values of $K_+$ and $K_-$ (all couplings are measured in units of $K_{LR}$, parameters with the dimension of an energy are measured in units of $K_{LR}^2$). Both step-like features are most pronounced in the
case of $K_\pm=0$, which is the case of only spin-flip tunneling term present in the original Hamiltonian. A finite $K_\pm$  corresponds to additional transversal coupling of the impurity spins,
which induces spin precession in the free case and is a source of independent spin-flips in the coupled case. This effect washes out the steps in the $I-V$ characteristics. Interestingly, only one
of $K_\pm$ being finite seems to cause a significantly weaker ``dephasing'' than a situation of both of them being non-zero.

\begin{center}
 \begin{figure}
 \includegraphics[width=\columnwidth]{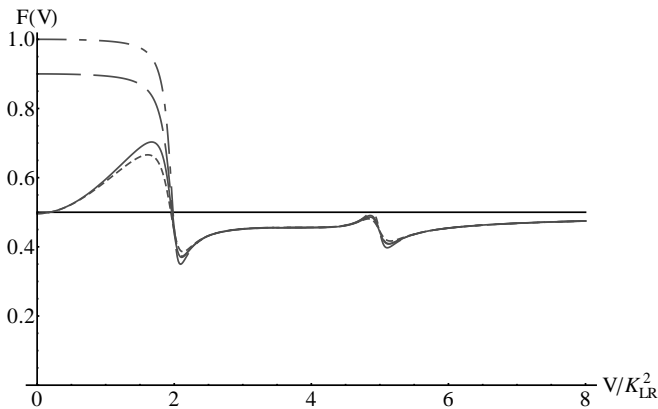}
 \caption{Fano factor for fixed magnetic fields
 $\Delta_1 = 2$ and $\Delta_2 = 5$ as a function of bias voltage. The different graphs correspond to $K_+ = 0 = K_-$ (dash-dotted line), $K_+ = 0.5 = K_-$ (short dashed line),
 $K_+ = 0.5$ and $K_- = 0$ (solid line) and $K_+ = 0$ and $K_- = 0.5$ (long dashed line). Solid line $F(V)=0.5$ is guide to the eye only.}
 \label{gra:fanosamedelta}
 \end{figure}
\end{center}

The Fano factor $F(V)$ as defined in Eq.~(\ref{eqn:fanofactor}) and $I(V)$ show up similar features at $V=\Delta_{1,2}$, see Fig.~\ref{gra:fanosamedelta}.

For small voltages and only spin-flip processes present $K_\pm=0$ the Fano factor approaches unity. For finite transversal coupling $K_\pm \neq 0$ this value becomes
nonuniversal. Interestingly, the effect of finite $K_+$ is much more pronounced than that of $K_-$. The reason for that is the fact that while finite $K_+$ indicates
the presence of the transversal couplings to the individual terminals, $K_-$ measures its asymmetry.

On the contrary, at $V \rightarrow \infty$ the Fano factor reaches the universal asymptotic value $1/2$, whatever the coupling strengths and temperature. This effect is usually observed
in the transport through constrictions with internal degrees of freedom, e.~g. it is known to appear in the Fano factor of the resonant level setup.\cite{AndersonFCS}

\subsection{Finite temperature effects}



To access the Johnson-Nyquist noise we set $V=0$ and assume $T$ to be small.\cite{Johnson1928,Nyquist1928} In this case the FCS reads

\begin{multline*}
 \ln \chi (\lambda) = \int\limits_0^\infty \frac{\text{d} \omega}{2 \pi} \ln \Big[\ \! 1 + n_F (1 - n_F) \\ \times \! \big[\ \! T_2(\omega)
 (e^{2 i \lambda} + e^{-2 i \lambda} - 2) + 2 T_1(\omega)
 (e^{i \lambda} + e^{-i \lambda} - 2) \big]\ \! \! \Big]\ .
\end{multline*}

\noindent Now we use the fact that $n_F(1-n_F) = - \beta^{-1} \partial_\omega n_F$ ($\beta = 1/T$ is the inverse temperature) and calculate the noise. Partial integration of the resulting expression gives

\begin{equation}
 S \approx 4 \beta^{-1} \frac{1}{2} [ T_1 (0) + 2 T_2(0) ] = 4 \beta^{-1} T_e \, ,
\end{equation}

\noindent where we identify $T_1 (0) + 2 T_2(0)$ with the effective transmission coefficient from
Section \ref{linearresponseregime}.



\begin{figure}
  \begin{overpic}[width=\columnwidth]{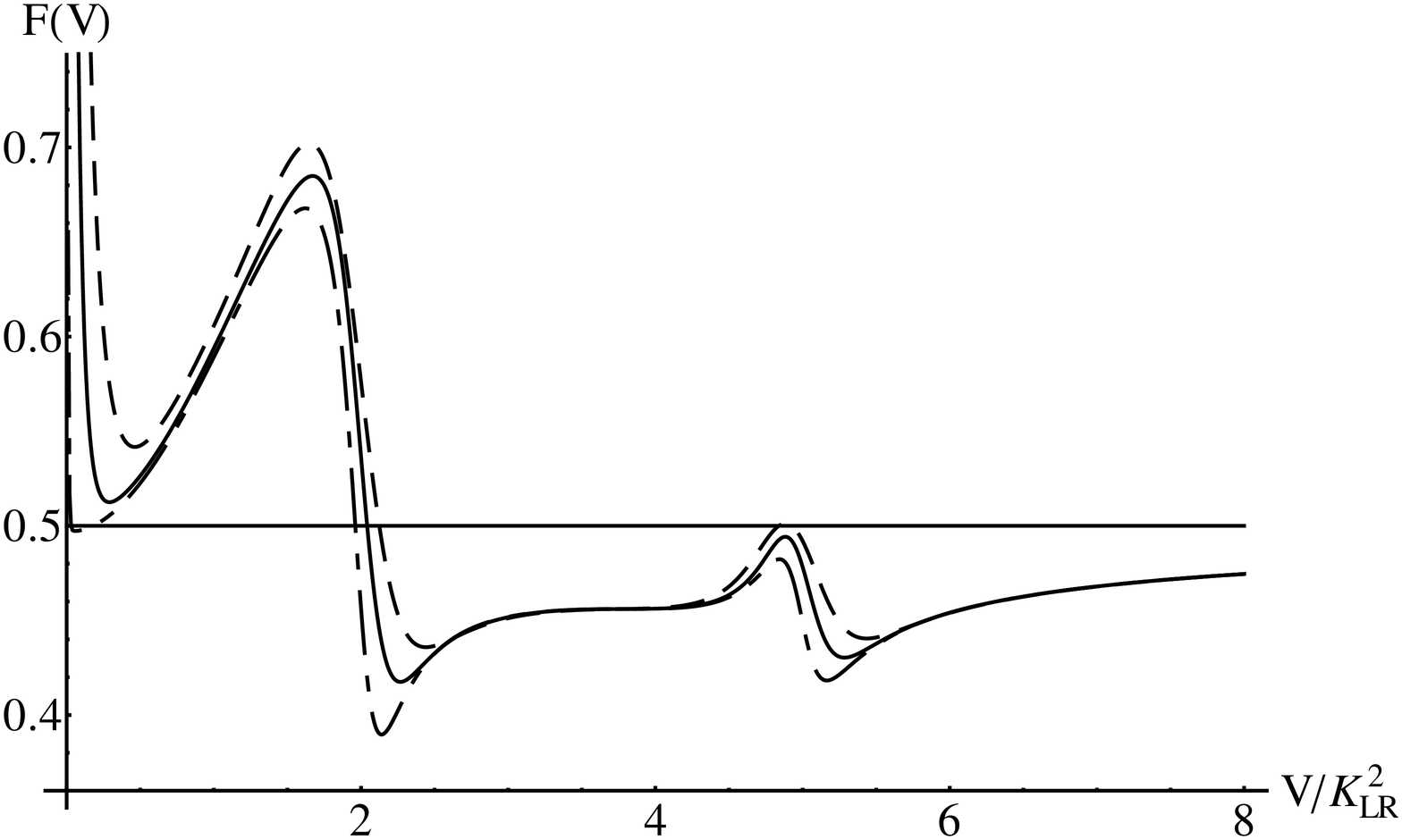}
    \put(35,25){\includegraphics[width=0.5\columnwidth]{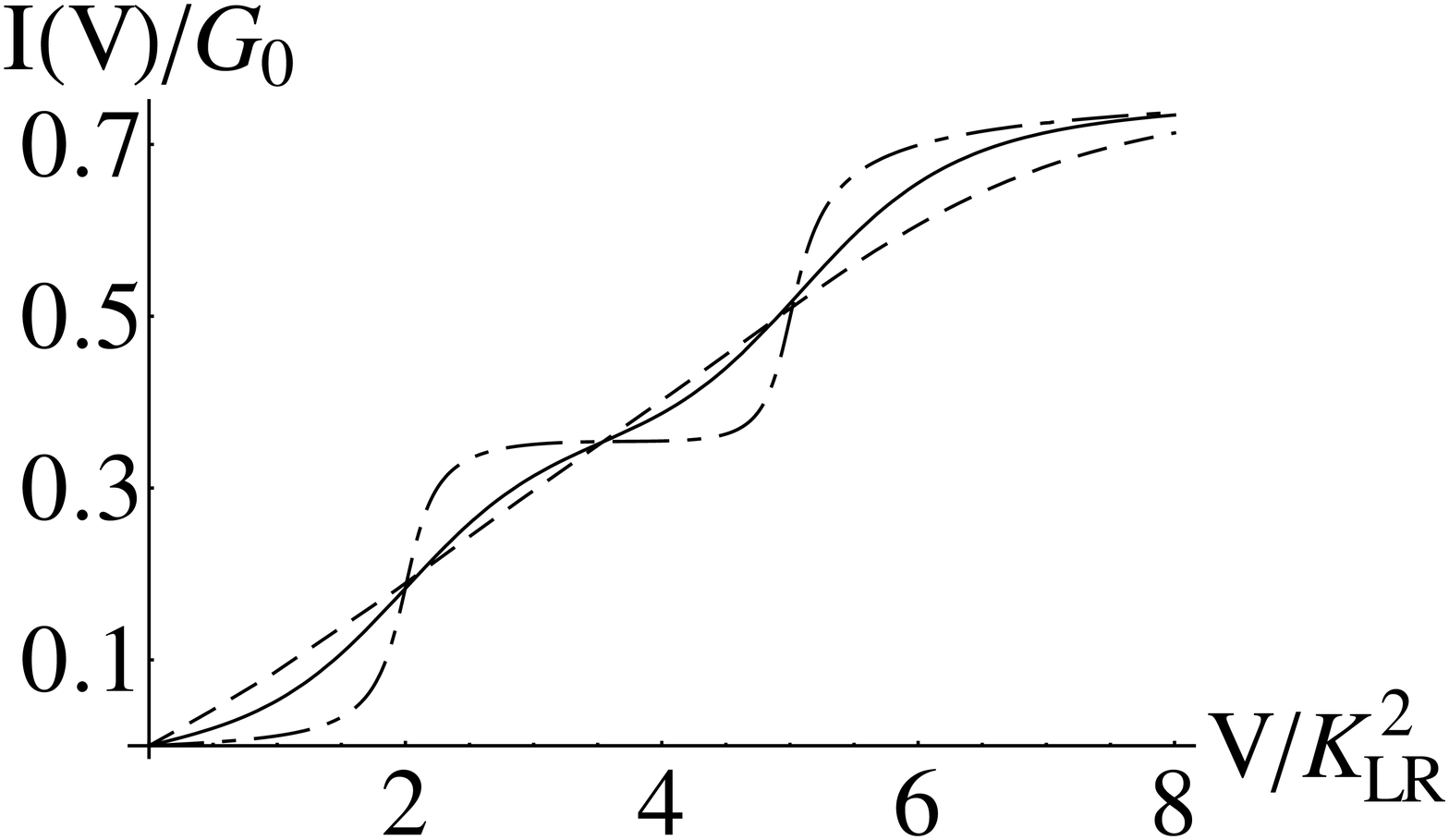}}
  \end{overpic}
\caption{{\bf Main graph:} Fano factor as a function of bias voltage for different temperatures and fixed magnetic fields $\Delta_1 = 2$ and $\Delta_2=5$, $T=0.005$ (dash-dotted line), $T=0.05$ (solid line), $T=0.1$ (dashed line).
The divergence of the Fano factor for $V \to 0$ can be explained by expanding noise and current (see text).
{\bf Inset:} Current as a function of bias voltage at different temperatures: $T=0.005$ (dash-dotted line), $T=0.5$ (solid line) and $T=1.$ (dashed line). Solid line $F(V)=0.5$ is guide to the eye only.}
\label{gra:fanotemp}
\end{figure}
In Fig.~\ref{gra:fanotemp} we show the behaviour of the Fano factor as a function of voltage for different temperatures. As expected the Fano factor grows for
increasing temperatures due to the onset of thermal fluctuations. The temperatures chosen are much smaller than $K_{LR}$ which represents the Kondo temperature because
any $T$ comparable or even higher than $K_{LR}$ eliminates all features from the plot as already can be seen from the inset. In addition to the features already
discussed in the $T=0$ case the Fano factor rises for $V \to 0$. This behaviour can be explained by the fact that the expansion of the current in $V$ starts with a
second order term while the noise starts with a first order term.

\section{Conclusions}

To conclude,
we have discussed the nonlinear transport properties of a double quantum dot in the Kondo regime. Using a series of dedicated transformations we rewrote the original Hamiltonian in one special region of the parameter space in terms of a quadratic Majorana resonant level model, which is conveniently diagonalizable even under nonequilibrium conditions. We have explicitly calculated the generating function of the full counting statistics and discussed its most prominent features, which reveal themselves in individual cumulants of the charge transport. In particular, we find a full suppression of transport for a special constellation of applied magnetic fields. We argue that this antiresonance feature is robust and universal even beyond the Toulouse point by an explicit perturbative expansion around it. One possible route for further progress would be an analysis of this phenomenon by  numerical and possibly more advanced analytical methods for the generic system parameters (e.~g. by quantum Monte Carlo methods or functional renormalization group approaches).

\acknowledgments The financial support was provided
by the DFG under grant No.~KO~2235/3, by the Kompetenznetz
``Funktionelle Nanostrukturen III'' of the
Baden-W\"urttemberg Stiftung, CQD and ``Enable fund'' of the University of Heidelberg.

\section*{Appendix}

Here we want to summarize the most important details of the perturbative expansion around the Toulouse point. The full fledged calculation of the corresponding $I(V)$ turns out to be very complex so that we employ the following approximation strategy. From the Toulouse point calculation we know that (at least at zero temperature) the electric current through the system is given by an energy integral over the voltage window of the imaginary part of the QD retarded GF
\begin{equation}
D_{a_+ a_+}^{\rm R}(t,t') = - i \Theta(t-t') \langle \{ a_+(t),  a_+ (t^\prime) \} \rangle \, ,
\end{equation}
which plays the role of the effective transmission coefficient (of course, this is compatible to the Wingreen-Meir formula\cite{Meir1992}). So we just need corrections to this GF. It is most conveniently done in the Keldysh formalism from the following GFs,
\begin{multline}
 D_{a_+ a_+}^{-- / -+} = - i \langle T_C \, a_+(t) a_+ (t^\prime)   \exp \Big\{- \gamma \int dt_1 \\  \big[ :\psi_s^\dagger (t_1) \psi_s (t_1):  \left[ a_+ (t_1) b_+ (t_1) + a_- (t_1) b_- (t_1) \right] \big] \Big\} \rangle \, ,
\end{multline}
which we expand for small $\gamma$. The first order vanishes due to the normal ordering in the free fermion sector. The second order we find to be given by
\begin{multline}
 \delta D_{a_+ a_+}^{-- / -+} = - \frac{ i \gamma^2}{2} \int dt_1 dt_2 \langle T_C \psi_1^\dagger \psi_1 \psi_2^\dagger \psi_2 \rangle \\
 \langle T_C a_+(t) a_+ (t^\prime) \left[ a_{+1} b_{+1} + a_{-1} b_{-1} \right] \left[ a_{+2} b_{+2} + a_{-2} b_{-2} \right] \rangle, \notag
 \label{eqn:correction}
\end{multline}
where the second index is the label of the time argument. Schematically it corresponds to a simple compound Majorana--free fermion loop diagram.\cite{Majumdar1998} Next we use Wick's theorem and find the first expectation value of the free fermion operators to be the product
of two GF. The Majorana part is a bit more involved. Since many contributions vanish for $\Delta_1 = - \Delta_2$ one finds a manageable expression
\begin{eqnarray}
 -i D_{a_+ a_+}(t-t_1) \big[ D_{b_- a_+}(t_2 -t^\prime) D_{b_+ a_-}(t_1 - t_2) \\ - D_{b_+ b_+}(t_1 - t_2) D_{a_+ a_+}(t_2 - t') \big] \\
  +i D_{a_+ b_-}(t - t_1) \big[ D_{b_- a_+} (t_2 - t^\prime) D_{a_- a_-} (t_1 - t_2) \\ - D_{a_+ a_+} (t_2 - t^\prime) D_{a_- b_+}(t_1 - t_2) \big] \, ,
  \label{eqn:secondorder}
\end{eqnarray}
where all GFs are to be understood as matrices in the Keldysh space.  Transformed into Fourier space we are left with expressions of the type
\begin{equation}
 D_1(\omega) \int d \epsilon D_2(\epsilon) \int d \Omega G_{\psi}(\omega - \epsilon + \Omega) G_{\psi}(\Omega) D_3(\omega) \, ,
 \label{eqn:fourierintegral}
\end{equation}
where $G_\psi$ are the (local, taken at $x=0$) free fermion GFs of $\psi_s$-fields, which are known, see e.~g. Ref.~[\onlinecite{AndersonFCS}]. $D_{1,2}$ are the ``outer'' GFs from Eq.~(\ref{eqn:secondorder}), which correspond to the one with the time arguments $t-t_1$ and $t_2-t'$ in the time domain, while $D_2$ represents the ``inner'' GF with the argument $t_1-t_2$.

The calculation of the dot GFs is somewhat lengthy but straightforward by writing down the action

\begin{multline}
 S = S \lbrack \eta , \xi \rbrack + \int \frac{d\omega}{2 \pi} \Big[ \sum\limits_{j= \pm} \alpha^T_j d_{ab}^{-1}(\Delta_+,\omega) \alpha_j \\
 - i \Delta_- (a_- b_+ + a_+ b_-) + i K_+ \eta_{sf} b_+ + i K_{LR} \xi_f a_+ \Big],
\end{multline}

\noindent where we defined $\Delta_\pm = \Delta_1 \pm \Delta_2$ and the superfield $\alpha_i^T = (a_{i-},a_{i+},b_{i-},b_{i+})$, and integrating out the lead fermions. $d_{ab}(\Delta_+,\omega)$ denotes the $4\times 4$-matrix GF for the constriction without couplings to the electrodes.
Integrating them out we can read off the necessary GFs from the GF for the superfields which has the structure:
\begin{equation}
 {\bf D} = \begin{pmatrix}
            D_{a_+ a_+} & \hat{D}_{a_+ b_+} & \hat{D}_{a_+ a_-} & D_{a_+ b_-} \\
	    \hat{D}_{b_+ a_+} & D_{b_+ b_+} & D_{b_+ a_-} & \hat{D}_{b_+ b_-} \\
	    \hat{D}_{a_- a_+} & D_{a_- b_+} & D_{a_- a_-} & \hat{D}_{a_- b_-} \\
	    D_{b_- a_+} & \hat{D}_{b_- b_+} & \hat{D}_{b_- a_-} & D_{b_- b_-}
           \end{pmatrix} \, ,
\end{equation}

\noindent where we indicated those components that are zero with a hat. Next we use the above GF and expand the structures \eqref{eqn:fourierintegral} to the lowest order in $\omega$. In total one needs to take care of
16 different terms of the type \eqref{eqn:fourierintegral} for both $--$ and $++$ component at $\omega \rightarrow 0$. All but two terms yield higher order contributions to the expansion. The only term in the $--$ component is the expression
\begin{equation}
 - \frac{i}{\Delta_-^2} \int d \epsilon \int d \Omega \, D^{--}_{a_- a_-} (\epsilon) G^{--}(\Omega-\epsilon)G^{--}(\Omega) \, .
\end{equation}
Performing the $\Omega$-integration first leaves us with an integral over $D^{--}_{a_- a_-}(\epsilon)$. This GF is an odd function in $\epsilon$ and has to be multiplied
with a function that
includes the bandwidth but is even in $\epsilon$. This means that the integral vanishes since integration and taking the limit for the bandwidth can be exchanged.

The only term that is non-zero in the lowest order in $\omega$ is given by

\begin{multline}
 i D^{--}_{a_+ b_-} D^{++}_{b_-a_+} \int d \epsilon D^{-+}_{a_- a_-} (\epsilon) \int d \Omega G^{-+}(\Omega-\epsilon)G^{+-}(\Omega) \\
 = \frac{K_{LR}^2}{\Delta_-^2} \int\limits_{-V}^{0} \frac{\text{d}\epsilon \; \epsilon^3}{(\Delta_-^2 - \epsilon^2)^2 + K_{LR}^4 \epsilon^2} \\
 = \text{const} \, \, V^4 + O(V^5).
\end{multline}
That means that this contribution vanishes for small voltages much faster than the leading order term and therefore ensures the existence of the perfect antiresonance in vicinity of the Toulouse point.

\bibliography{DD_Kondo_paper2}

\begin{thebibliography}{54}
\expandafter\ifx\csname natexlab\endcsname\relax\def\natexlab#1{#1}\fi
\expandafter\ifx\csname bibnamefont\endcsname\relax
  \def\bibnamefont#1{#1}\fi
\expandafter\ifx\csname bibfnamefont\endcsname\relax
  \def\bibfnamefont#1{#1}\fi
\expandafter\ifx\csname citenamefont\endcsname\relax
  \def\citenamefont#1{#1}\fi
\expandafter\ifx\csname url\endcsname\relax
  \def\url#1{\texttt{#1}}\fi
\expandafter\ifx\csname urlprefix\endcsname\relax\def\urlprefix{URL }\fi
\providecommand{\bibinfo}[2]{#2}
\providecommand{\eprint}[2][]{\url{#2}}

\bibitem[{\citenamefont{Nazarov and Blanter}(2009)}]{Nazarov2009}
\bibinfo{author}{\bibfnamefont{Y.~V.} \bibnamefont{Nazarov}} \bibnamefont{and}
  \bibinfo{author}{\bibfnamefont{Y.~M.} \bibnamefont{Blanter}},
  \emph{\bibinfo{title}{Quantum Transport: Introduction to Nanoscience}}
  (\bibinfo{publisher}{Cambridge University Press}, \bibinfo{year}{2009}).

\bibitem[{\citenamefont{Sela and Affleck}(2009{\natexlab{a}})}]{sela:125110}
\bibinfo{author}{\bibfnamefont{E.}~\bibnamefont{Sela}} \bibnamefont{and}
  \bibinfo{author}{\bibfnamefont{I.}~\bibnamefont{Affleck}},
  \bibinfo{journal}{Phys. Rev. B} \textbf{\bibinfo{volume}{79}},
  \bibinfo{eid}{125110} (\bibinfo{year}{2009}{\natexlab{a}}).

\bibitem[{\citenamefont{Gan}(1995)}]{PhysRevB.51.8287}
\bibinfo{author}{\bibfnamefont{J.}~\bibnamefont{Gan}}, \bibinfo{journal}{Phys.
  Rev. B} \textbf{\bibinfo{volume}{51}}, \bibinfo{pages}{8287}
  (\bibinfo{year}{1995}).

\bibitem[{\citenamefont{M{\"{u}}ller et~al.}(2010)\citenamefont{M{\"{u}}ller,
  Koerting, Schuricht, and Andergassen}}]{0295-5075-92-1-10002}
\bibinfo{author}{\bibfnamefont{S.~Y.} \bibnamefont{M{\"{u}}ller}},
  \bibinfo{author}{\bibfnamefont{V.}~\bibnamefont{Koerting}},
  \bibinfo{author}{\bibfnamefont{D.}~\bibnamefont{Schuricht}},
  \bibnamefont{and}
  \bibinfo{author}{\bibfnamefont{S.}~\bibnamefont{Andergassen}},
  \bibinfo{journal}{Europhys. Lett.} \textbf{\bibinfo{volume}{92}},
  \bibinfo{pages}{10002} (\bibinfo{year}{2010}).

\bibitem[{\citenamefont{Koerting et~al.}(2007)\citenamefont{Koerting, W\"olfle,
  and Paaske}}]{PhysRevLett.99.036807}
\bibinfo{author}{\bibfnamefont{V.}~\bibnamefont{Koerting}},
  \bibinfo{author}{\bibfnamefont{P.}~\bibnamefont{W\"olfle}}, \bibnamefont{and}
  \bibinfo{author}{\bibfnamefont{J.}~\bibnamefont{Paaske}},
  \bibinfo{journal}{Phys. Rev. Lett.} \textbf{\bibinfo{volume}{99}},
  \bibinfo{pages}{036807} (\bibinfo{year}{2007}).

\bibitem[{\citenamefont{van~der Wiel et~al.}(2002)\citenamefont{van~der Wiel,
  De~Franceschi, Elzerman, Fujisawa, Tarucha, and
  Kouwenhoven}}]{RevModPhys.75.1}
\bibinfo{author}{\bibfnamefont{W.~G.} \bibnamefont{van~der Wiel}},
  \bibinfo{author}{\bibfnamefont{S.}~\bibnamefont{De~Franceschi}},
  \bibinfo{author}{\bibfnamefont{J.~M.} \bibnamefont{Elzerman}},
  \bibinfo{author}{\bibfnamefont{T.}~\bibnamefont{Fujisawa}},
  \bibinfo{author}{\bibfnamefont{S.}~\bibnamefont{Tarucha}}, \bibnamefont{and}
  \bibinfo{author}{\bibfnamefont{L.~P.} \bibnamefont{Kouwenhoven}},
  \bibinfo{journal}{Rev. Mod. Phys.} \textbf{\bibinfo{volume}{75}},
  \bibinfo{pages}{1} (\bibinfo{year}{2002}).

\bibitem[{\citenamefont{Sztenkiel and
  Swirkowicz}(2007)}]{0953-8984-19-25-256205}
\bibinfo{author}{\bibfnamefont{D.}~\bibnamefont{Sztenkiel}} \bibnamefont{and}
  \bibinfo{author}{\bibfnamefont{R.}~\bibnamefont{Swirkowicz}},
  \bibinfo{journal}{Journal of Physics: Condensed Matter}
  \textbf{\bibinfo{volume}{19}}, \bibinfo{pages}{256205}
  (\bibinfo{year}{2007}).

\bibitem[{\citenamefont{{Teemu Pohjola} et~al.}(2001)\citenamefont{{Teemu
  Pohjola}, {Herbert Schoeller}, and {Gerd Sch\"on}}}]{refId}
\bibinfo{author}{\bibnamefont{{Teemu Pohjola}}},
  \bibinfo{author}{\bibnamefont{{Herbert Schoeller}}}, \bibnamefont{and}
  \bibinfo{author}{\bibnamefont{{Gerd Sch\"on}}}, \bibinfo{journal}{Europhys.
  Lett.} \textbf{\bibinfo{volume}{54}}, \bibinfo{pages}{241}
  (\bibinfo{year}{2001}).

\bibitem[{\citenamefont{da~Silva et~al.}(2008)\citenamefont{da~Silva, Sandler,
  Ingersent, and Ulloa}}]{DiasdaSilva20081002}
\bibinfo{author}{\bibfnamefont{L.~G.~D.} \bibnamefont{da~Silva}},
  \bibinfo{author}{\bibfnamefont{N.}~\bibnamefont{Sandler}},
  \bibinfo{author}{\bibfnamefont{K.}~\bibnamefont{Ingersent}},
  \bibnamefont{and} \bibinfo{author}{\bibfnamefont{S.~E.} \bibnamefont{Ulloa}},
  \bibinfo{journal}{Physica E: Low-dimensional Systems and Nanostructures}
  \textbf{\bibinfo{volume}{40}}, \bibinfo{pages}{1002 } (\bibinfo{year}{2008}),
  \bibinfo{note}{17th International Conference on Electronic Properties of
  Two-Dimensional Systems}.

\bibitem[{\citenamefont{Andergassen et~al.}(2008)\citenamefont{Andergassen,
  Simon, Florens, and Feinberg}}]{PhysRevB.77.045309}
\bibinfo{author}{\bibfnamefont{S.}~\bibnamefont{Andergassen}},
  \bibinfo{author}{\bibfnamefont{P.}~\bibnamefont{Simon}},
  \bibinfo{author}{\bibfnamefont{S.}~\bibnamefont{Florens}}, \bibnamefont{and}
  \bibinfo{author}{\bibfnamefont{D.}~\bibnamefont{Feinberg}},
  \bibinfo{journal}{Phys. Rev. B} \textbf{\bibinfo{volume}{77}},
  \bibinfo{pages}{045309} (\bibinfo{year}{2008}).

\bibitem[{\citenamefont{Dias~da Silva et~al.}(2008)\citenamefont{Dias~da Silva,
  Ingersent, Sandler, and Ulloa}}]{PhysRevB.78.153304}
\bibinfo{author}{\bibfnamefont{L.~G. G.~V.} \bibnamefont{Dias~da Silva}},
  \bibinfo{author}{\bibfnamefont{K.}~\bibnamefont{Ingersent}},
  \bibinfo{author}{\bibfnamefont{N.}~\bibnamefont{Sandler}}, \bibnamefont{and}
  \bibinfo{author}{\bibfnamefont{S.~E.} \bibnamefont{Ulloa}},
  \bibinfo{journal}{Phys. Rev. B} \textbf{\bibinfo{volume}{78}},
  \bibinfo{pages}{153304} (\bibinfo{year}{2008}).

\bibitem[{\citenamefont{Fang and Luo}(2010)}]{PhysRevB.81.113402}
\bibinfo{author}{\bibfnamefont{T.-F.} \bibnamefont{Fang}} \bibnamefont{and}
  \bibinfo{author}{\bibfnamefont{H.-G.} \bibnamefont{Luo}},
  \bibinfo{journal}{Phys. Rev. B} \textbf{\bibinfo{volume}{81}},
  \bibinfo{pages}{113402} (\bibinfo{year}{2010}).

\bibitem[{\citenamefont{Simon and Feinberg}(2006)}]{PhysRevLett.97.247207}
\bibinfo{author}{\bibfnamefont{P.}~\bibnamefont{Simon}} \bibnamefont{and}
  \bibinfo{author}{\bibfnamefont{D.}~\bibnamefont{Feinberg}},
  \bibinfo{journal}{Phys. Rev. Lett.} \textbf{\bibinfo{volume}{97}},
  \bibinfo{pages}{247207} (\bibinfo{year}{2006}).

\bibitem[{\citenamefont{Sela et~al.}(2011)\citenamefont{Sela, Mitchell, and
  Fritz}}]{PhysRevLett.106.147202}
\bibinfo{author}{\bibfnamefont{E.}~\bibnamefont{Sela}},
  \bibinfo{author}{\bibfnamefont{A.~K.} \bibnamefont{Mitchell}},
  \bibnamefont{and} \bibinfo{author}{\bibfnamefont{L.}~\bibnamefont{Fritz}},
  \bibinfo{journal}{Phys. Rev. Lett.} \textbf{\bibinfo{volume}{106}},
  \bibinfo{pages}{147202} (\bibinfo{year}{2011}).

\bibitem[{\citenamefont{Sela and Affleck}(2009{\natexlab{b}})}]{Sela2009}
\bibinfo{author}{\bibfnamefont{E.}~\bibnamefont{Sela}} \bibnamefont{and}
  \bibinfo{author}{\bibfnamefont{I.}~\bibnamefont{Affleck}},
  \bibinfo{journal}{Phys. Rev. Lett.} \textbf{\bibinfo{volume}{102}},
  \bibinfo{pages}{047201} (\bibinfo{year}{2009}{\natexlab{b}}).

\bibitem[{\citenamefont{Anderson}(1961)}]{Anderson1961}
\bibinfo{author}{\bibfnamefont{P.~W.} \bibnamefont{Anderson}},
  \bibinfo{journal}{Phys. Rev.} \textbf{\bibinfo{volume}{124}},
  \bibinfo{pages}{41} (\bibinfo{year}{1961}).

\bibitem[{\citenamefont{Hewson}(1997)}]{Hewson1997}
\bibinfo{author}{\bibfnamefont{A.~C.} \bibnamefont{Hewson}},
  \emph{\bibinfo{title}{The Kondo Problem to Heavy Fermions}}
  (\bibinfo{publisher}{Cambridge University Press, Cambridge},
  \bibinfo{year}{1997}).

\bibitem[{\citenamefont{Konik}(2007)}]{PhysRevLett.99.076602}
\bibinfo{author}{\bibfnamefont{R.~M.} \bibnamefont{Konik}},
  \bibinfo{journal}{Phys. Rev. Lett.} \textbf{\bibinfo{volume}{99}},
  \bibinfo{pages}{076602} (\bibinfo{year}{2007}).

\bibitem[{\citenamefont{Meden and Marquardt}(2006)}]{PhysRevLett.96.146801}
\bibinfo{author}{\bibfnamefont{V.}~\bibnamefont{Meden}} \bibnamefont{and}
  \bibinfo{author}{\bibfnamefont{F.}~\bibnamefont{Marquardt}},
  \bibinfo{journal}{Phys. Rev. Lett.} \textbf{\bibinfo{volume}{96}},
  \bibinfo{pages}{146801} (\bibinfo{year}{2006}).

\bibitem[{\citenamefont{Kubo et~al.}(2011)\citenamefont{Kubo, Tokura, and
  Tarucha}}]{PhysRevB.83.115310}
\bibinfo{author}{\bibfnamefont{T.}~\bibnamefont{Kubo}},
  \bibinfo{author}{\bibfnamefont{Y.}~\bibnamefont{Tokura}}, \bibnamefont{and}
  \bibinfo{author}{\bibfnamefont{S.}~\bibnamefont{Tarucha}},
  \bibinfo{journal}{Phys. Rev. B} \textbf{\bibinfo{volume}{83}},
  \bibinfo{pages}{115310} (\bibinfo{year}{2011}).

\bibitem[{\citenamefont{Glazman}(2000)}]{GlazmanAnkara}
\bibinfo{author}{\bibfnamefont{L.~I.} \bibnamefont{Glazman}}, in
  \emph{\bibinfo{booktitle}{Quantum Mesoscopic Phenomena and Mesoscopic Devices
  in Microelectronics}}, edited by \bibinfo{editor}{\bibfnamefont{I.~O.}
  \bibnamefont{Kulik}} \bibnamefont{and}
  \bibinfo{editor}{\bibfnamefont{R.}~\bibnamefont{Ellialtioglu}}
  (\bibinfo{publisher}{Kluwer Academic Publishers},
  \bibinfo{address}{Dordrecht/Boston/London}, \bibinfo{year}{2000}), NATO ASI.

\bibitem[{\citenamefont{Meir and Golub}(2002)}]{Meir2002}
\bibinfo{author}{\bibfnamefont{Y.}~\bibnamefont{Meir}} \bibnamefont{and}
  \bibinfo{author}{\bibfnamefont{A.}~\bibnamefont{Golub}},
  \bibinfo{journal}{Phys. Rev. Lett.} \textbf{\bibinfo{volume}{88}},
  \bibinfo{pages}{116802} (\bibinfo{year}{2002}).

\bibitem[{\citenamefont{Sela et~al.}(2006)\citenamefont{Sela, Oreg, {von
  Oppen}, and Koch}}]{Sela2006}
\bibinfo{author}{\bibfnamefont{E.}~\bibnamefont{Sela}},
  \bibinfo{author}{\bibfnamefont{Y.}~\bibnamefont{Oreg}},
  \bibinfo{author}{\bibfnamefont{F.}~\bibnamefont{{von Oppen}}},
  \bibnamefont{and} \bibinfo{author}{\bibfnamefont{J.}~\bibnamefont{Koch}},
  \bibinfo{journal}{Phys. Rev. Lett.} \textbf{\bibinfo{volume}{97}},
  \bibinfo{eid}{086601} (\bibinfo{year}{2006}).

\bibitem[{\citenamefont{Golub}(2006)}]{Golub2006}
\bibinfo{author}{\bibfnamefont{A.}~\bibnamefont{Golub}},
  \bibinfo{journal}{Phys. Rev. B} \textbf{\bibinfo{volume}{73}},
  \bibinfo{eid}{233310} (\bibinfo{year}{2006}).

\bibitem[{\citenamefont{Gogolin and Komnik}(2006{\natexlab{a}})}]{Gogolin2006}
\bibinfo{author}{\bibfnamefont{A.~O.} \bibnamefont{Gogolin}} \bibnamefont{and}
  \bibinfo{author}{\bibfnamefont{A.}~\bibnamefont{Komnik}},
  \bibinfo{journal}{Phys. Rev. Lett.} \textbf{\bibinfo{volume}{97}},
  \bibinfo{eid}{016602} (\bibinfo{year}{2006}{\natexlab{a}}).

\bibitem[{\citenamefont{Gogolin and Komnik}(2006{\natexlab{b}})}]{AndersonFCS}
\bibinfo{author}{\bibfnamefont{A.~O.} \bibnamefont{Gogolin}} \bibnamefont{and}
  \bibinfo{author}{\bibfnamefont{A.}~\bibnamefont{Komnik}},
  \bibinfo{journal}{Phys. Rev. B} \textbf{\bibinfo{volume}{73}},
  \bibinfo{pages}{195301} (\bibinfo{year}{2006}{\natexlab{b}}).

\bibitem[{\citenamefont{Zarchin et~al.}(2008)\citenamefont{Zarchin, Zaffalon,
  Heiblum, Mahalu, and Umansky}}]{Zarchin2008}
\bibinfo{author}{\bibfnamefont{O.}~\bibnamefont{Zarchin}},
  \bibinfo{author}{\bibfnamefont{M.}~\bibnamefont{Zaffalon}},
  \bibinfo{author}{\bibfnamefont{M.}~\bibnamefont{Heiblum}},
  \bibinfo{author}{\bibfnamefont{D.}~\bibnamefont{Mahalu}}, \bibnamefont{and}
  \bibinfo{author}{\bibfnamefont{V.}~\bibnamefont{Umansky}},
  \bibinfo{journal}{Phys. Rev. B} \textbf{\bibinfo{volume}{77}},
  \bibinfo{eid}{241303} (\bibinfo{year}{2008}).

\bibitem[{\citenamefont{Toulouse}(1969)}]{Toulouse1969a}
\bibinfo{author}{\bibfnamefont{G.}~\bibnamefont{Toulouse}},
  \bibinfo{journal}{C. R. Acad. Sci. Ser. B} \textbf{\bibinfo{volume}{268}},
  \bibinfo{pages}{1200} (\bibinfo{year}{1969}).

\bibitem[{\citenamefont{Emery and Kivelson}(1992)}]{Emery1992}
\bibinfo{author}{\bibfnamefont{V.~J.} \bibnamefont{Emery}} \bibnamefont{and}
  \bibinfo{author}{\bibfnamefont{S.}~\bibnamefont{Kivelson}},
  \bibinfo{journal}{Phys. Rev. B} \textbf{\bibinfo{volume}{46}},
  \bibinfo{pages}{10812} (\bibinfo{year}{1992}).

\bibitem[{\citenamefont{Schiller and Hershfield}(1998)}]{Schiller1998}
\bibinfo{author}{\bibfnamefont{A.}~\bibnamefont{Schiller}} \bibnamefont{and}
  \bibinfo{author}{\bibfnamefont{S.}~\bibnamefont{Hershfield}},
  \bibinfo{journal}{Phys. Rev. B} \textbf{\bibinfo{volume}{58}},
  \bibinfo{pages}{14978} (\bibinfo{year}{1998}).

\bibitem[{\citenamefont{Gogolin et~al.}(1998)\citenamefont{Gogolin, Nersesyan,
  and Tsvelik}}]{Gogolin1998}
\bibinfo{author}{\bibfnamefont{A.~O.} \bibnamefont{Gogolin}},
  \bibinfo{author}{\bibfnamefont{A.~A.} \bibnamefont{Nersesyan}},
  \bibnamefont{and} \bibinfo{author}{\bibfnamefont{A.~M.}
  \bibnamefont{Tsvelik}}, \emph{\bibinfo{title}{Bosonization and Strongly
  Correlated Systems}} (\bibinfo{publisher}{Cambridge University Press},
  \bibinfo{address}{Cambridge}, \bibinfo{year}{1998}).

\bibitem[{\citenamefont{\ifmmode~\check{Z}\else \v{Z}\fi{}itko and
  Bon\ifmmode~\check{c}\else \v{c}\fi{}a}(2006)}]{Zitko2006}
\bibinfo{author}{\bibfnamefont{R.}~\bibnamefont{\ifmmode~\check{Z}\else
  \v{Z}\fi{}itko}} \bibnamefont{and}
  \bibinfo{author}{\bibfnamefont{J.}~\bibnamefont{Bon\ifmmode~\check{c}\else
  \v{c}\fi{}a}}, \bibinfo{journal}{Phys. Rev. B} \textbf{\bibinfo{volume}{74}},
  \bibinfo{pages}{045312} (\bibinfo{year}{2006}).

\bibitem[{\citenamefont{Schrieffer and Wolff}(1966)}]{Schrieffer1966}
\bibinfo{author}{\bibfnamefont{J.~R.} \bibnamefont{Schrieffer}}
  \bibnamefont{and} \bibinfo{author}{\bibfnamefont{P.~A.} \bibnamefont{Wolff}},
  \bibinfo{journal}{Phys. Rev.} \textbf{\bibinfo{volume}{149}},
  \bibinfo{pages}{491} (\bibinfo{year}{1966}).

\bibitem[{\citenamefont{Giamarchi}(2004)}]{Giamarchi2004}
\bibinfo{author}{\bibfnamefont{T.}~\bibnamefont{Giamarchi}},
  \emph{\bibinfo{title}{Quantum physics in one dimension}}, International
  series of monographs on physics (\bibinfo{publisher}{Clarendon},
  \bibinfo{year}{2004}).

\bibitem[{\citenamefont{Kawaguchi}(2009)}]{Kawaguchi2009}
\bibinfo{author}{\bibfnamefont{S.}~\bibnamefont{Kawaguchi}},
  \bibinfo{journal}{J. Phys.: Condens. Matter} \textbf{\bibinfo{volume}{21}},
  \bibinfo{pages}{395303} (\bibinfo{year}{2009}).

\bibitem[{\citenamefont{Majumdar et~al.}(1998)\citenamefont{Majumdar, Schiller,
  and Hershfield}}]{Majumdar1998}
\bibinfo{author}{\bibfnamefont{K.}~\bibnamefont{Majumdar}},
  \bibinfo{author}{\bibfnamefont{A.}~\bibnamefont{Schiller}}, \bibnamefont{and}
  \bibinfo{author}{\bibfnamefont{S.}~\bibnamefont{Hershfield}},
  \bibinfo{journal}{Phys. Rev. B} \textbf{\bibinfo{volume}{57}},
  \bibinfo{pages}{2991} (\bibinfo{year}{1998}).

\bibitem[{\citenamefont{Levitov et~al.}(1996)\citenamefont{Levitov, Lee, and
  Lesovik}}]{lll}
\bibinfo{author}{\bibfnamefont{L.~S.} \bibnamefont{Levitov}},
  \bibinfo{author}{\bibfnamefont{W.~W.} \bibnamefont{Lee}}, \bibnamefont{and}
  \bibinfo{author}{\bibfnamefont{G.~B.} \bibnamefont{Lesovik}},
  \bibinfo{journal}{J. Math. Phys.} \textbf{\bibinfo{volume}{37}},
  \bibinfo{pages}{4845} (\bibinfo{year}{1996}).

\bibitem[{\citenamefont{Fano}(1947)}]{PhysRev.72.26}
\bibinfo{author}{\bibfnamefont{U.}~\bibnamefont{Fano}}, \bibinfo{journal}{Phys.
  Rev.} \textbf{\bibinfo{volume}{72}}, \bibinfo{pages}{26}
  (\bibinfo{year}{1947}).

\bibitem[{\citenamefont{Komnik and Gogolin}(2005)}]{Komnik2005}
\bibinfo{author}{\bibfnamefont{A.}~\bibnamefont{Komnik}} \bibnamefont{and}
  \bibinfo{author}{\bibfnamefont{A.~O.} \bibnamefont{Gogolin}},
  \bibinfo{journal}{Phys.~Rev.~Lett.} \textbf{\bibinfo{volume}{94}},
  \bibinfo{pages}{216601} (\bibinfo{year}{2005}).

\bibitem[{\citenamefont{Levitov and Reznikov}(2004)}]{levitov04}
\bibinfo{author}{\bibfnamefont{L.~S.} \bibnamefont{Levitov}} \bibnamefont{and}
  \bibinfo{author}{\bibfnamefont{M.}~\bibnamefont{Reznikov}},
  \bibinfo{journal}{Phys. Rev. B} \textbf{\bibinfo{volume}{70}},
  \bibinfo{pages}{115305} (\bibinfo{year}{2004}).

\bibitem[{\citenamefont{Levitov and Lesovik}(1993)}]{Levitov1993}
\bibinfo{author}{\bibfnamefont{L.~S.} \bibnamefont{Levitov}} \bibnamefont{and}
  \bibinfo{author}{\bibfnamefont{G.~B.} \bibnamefont{Lesovik}},
  \bibinfo{journal}{JETP Lett.} \textbf{\bibinfo{volume}{58}},
  \bibinfo{pages}{230} (\bibinfo{year}{1993}).

\bibitem[{\citenamefont{Feynman}(1939)}]{Feynman1939}
\bibinfo{author}{\bibfnamefont{R.~P.} \bibnamefont{Feynman}},
  \bibinfo{journal}{Phys.~Rev.} \textbf{\bibinfo{volume}{56}},
  \bibinfo{pages}{340} (\bibinfo{year}{1939}).

\bibitem[{\citenamefont{Schmidt et~al.}(2007)\citenamefont{Schmidt, Gogolin,
  and Komnik}}]{schmidt:235105}
\bibinfo{author}{\bibfnamefont{T.~L.} \bibnamefont{Schmidt}},
  \bibinfo{author}{\bibfnamefont{A.~O.} \bibnamefont{Gogolin}},
  \bibnamefont{and} \bibinfo{author}{\bibfnamefont{A.}~\bibnamefont{Komnik}},
  \bibinfo{journal}{Phys. Rev. B} \textbf{\bibinfo{volume}{75}},
  \bibinfo{eid}{235105} (\bibinfo{year}{2007}).

\bibitem[{\citenamefont{L{\'{o}}pez et~al.}(2005)\citenamefont{L{\'{o}}pez,
  S{\'{a}}nchez, Lee, Choi, Simon, and Le~Hur}}]{L'opez2005}
\bibinfo{author}{\bibfnamefont{R.}~\bibnamefont{L{\'{o}}pez}},
  \bibinfo{author}{\bibfnamefont{D.}~\bibnamefont{S{\'{a}}nchez}},
  \bibinfo{author}{\bibfnamefont{M.}~\bibnamefont{Lee}},
  \bibinfo{author}{\bibfnamefont{M.-S.} \bibnamefont{Choi}},
  \bibinfo{author}{\bibfnamefont{P.}~\bibnamefont{Simon}}, \bibnamefont{and}
  \bibinfo{author}{\bibfnamefont{K.}~\bibnamefont{Le~Hur}},
  \bibinfo{journal}{Phys. Rev. B} \textbf{\bibinfo{volume}{71}},
  \bibinfo{pages}{115312} (\bibinfo{year}{2005}).

\bibitem[{\citenamefont{Kashcheyevs et~al.}(2007)\citenamefont{Kashcheyevs,
  Schiller, Aharony, and Entin-Wohlman}}]{Kashcheyevs2007a}
\bibinfo{author}{\bibfnamefont{V.}~\bibnamefont{Kashcheyevs}},
  \bibinfo{author}{\bibfnamefont{A.}~\bibnamefont{Schiller}},
  \bibinfo{author}{\bibfnamefont{A.}~\bibnamefont{Aharony}}, \bibnamefont{and}
  \bibinfo{author}{\bibfnamefont{O.}~\bibnamefont{Entin-Wohlman}},
  \bibinfo{journal}{Phys. Rev. B} \textbf{\bibinfo{volume}{75}},
  \bibinfo{eid}{115313} (\bibinfo{year}{2007}).

\bibitem[{\citenamefont{Hackenbroich and
  Weidenm{\"{u}}ller}(1996)}]{Hackenbroich1996}
\bibinfo{author}{\bibfnamefont{G.}~\bibnamefont{Hackenbroich}}
  \bibnamefont{and} \bibinfo{author}{\bibfnamefont{H.~A.}
  \bibnamefont{Weidenm{\"{u}}ller}}, \bibinfo{journal}{Phys. Rev. Lett.}
  \textbf{\bibinfo{volume}{76}}, \bibinfo{pages}{110} (\bibinfo{year}{1996}).

\bibitem[{\citenamefont{Kubala and K{\"{o}}nig}(2003)}]{Kubala2003}
\bibinfo{author}{\bibfnamefont{B.}~\bibnamefont{Kubala}} \bibnamefont{and}
  \bibinfo{author}{\bibfnamefont{J.}~\bibnamefont{K{\"{o}}nig}},
  \bibinfo{journal}{Phys. Rev. B} \textbf{\bibinfo{volume}{67}},
  \bibinfo{pages}{205303} (\bibinfo{year}{2003}).

\bibitem[{\citenamefont{Dahlhaus et~al.}(2010)\citenamefont{Dahlhaus, Maier,
  and Komnik}}]{Dahlhaus2010}
\bibinfo{author}{\bibfnamefont{J.~P.} \bibnamefont{Dahlhaus}},
  \bibinfo{author}{\bibfnamefont{S.}~\bibnamefont{Maier}}, \bibnamefont{and}
  \bibinfo{author}{\bibfnamefont{A.}~\bibnamefont{Komnik}},
  \bibinfo{journal}{Phys. Rev. B} \textbf{\bibinfo{volume}{81}},
  \bibinfo{pages}{075110} (\bibinfo{year}{2010}).

\bibitem[{\citenamefont{Boese et~al.}(2002)\citenamefont{Boese, Hofstetter, and
  Schoeller}}]{PhysRevB.66.125315}
\bibinfo{author}{\bibfnamefont{D.}~\bibnamefont{Boese}},
  \bibinfo{author}{\bibfnamefont{W.}~\bibnamefont{Hofstetter}},
  \bibnamefont{and}
  \bibinfo{author}{\bibfnamefont{H.}~\bibnamefont{Schoeller}},
  \bibinfo{journal}{Phys. Rev. B} \textbf{\bibinfo{volume}{66}},
  \bibinfo{pages}{125315} (\bibinfo{year}{2002}).

\bibitem[{\citenamefont{Holleitner et~al.}(2001)\citenamefont{Holleitner,
  Decker, Qin, Eberl, and Blick}}]{Holleitner2001}
\bibinfo{author}{\bibfnamefont{A.~W.} \bibnamefont{Holleitner}},
  \bibinfo{author}{\bibfnamefont{C.~R.} \bibnamefont{Decker}},
  \bibinfo{author}{\bibfnamefont{H.}~\bibnamefont{Qin}},
  \bibinfo{author}{\bibfnamefont{K.}~\bibnamefont{Eberl}}, \bibnamefont{and}
  \bibinfo{author}{\bibfnamefont{R.~H.} \bibnamefont{Blick}},
  \bibinfo{journal}{Phys. Rev. Lett.} \textbf{\bibinfo{volume}{87}},
  \bibinfo{pages}{256802} (\bibinfo{year}{2001}).

\bibitem[{\citenamefont{Wilhelm et~al.}(2002)\citenamefont{Wilhelm, Schmid,
  Weis, and {von Klitzing}}}]{Wilhelm2002}
\bibinfo{author}{\bibfnamefont{U.}~\bibnamefont{Wilhelm}},
  \bibinfo{author}{\bibfnamefont{J.}~\bibnamefont{Schmid}},
  \bibinfo{author}{\bibfnamefont{J.}~\bibnamefont{Weis}}, \bibnamefont{and}
  \bibinfo{author}{\bibfnamefont{K.}~\bibnamefont{{von Klitzing}}},
  \bibinfo{journal}{Physica E} \textbf{\bibinfo{volume}{14}},
  \bibinfo{pages}{385 } (\bibinfo{year}{2002}).

\bibitem[{\citenamefont{Johnson}(1928)}]{Johnson1928}
\bibinfo{author}{\bibfnamefont{J.~B.} \bibnamefont{Johnson}},
  \bibinfo{journal}{Phys. Rev.} \textbf{\bibinfo{volume}{32}},
  \bibinfo{pages}{97} (\bibinfo{year}{1928}).

\bibitem[{\citenamefont{Nyquist}(1928)}]{Nyquist1928}
\bibinfo{author}{\bibfnamefont{H.}~\bibnamefont{Nyquist}},
  \bibinfo{journal}{Phys. Rev.} \textbf{\bibinfo{volume}{32}},
  \bibinfo{pages}{110} (\bibinfo{year}{1928}).

\bibitem[{\citenamefont{Meir and Wingreen}(1992)}]{Meir1992}
\bibinfo{author}{\bibfnamefont{Y.}~\bibnamefont{Meir}} \bibnamefont{and}
  \bibinfo{author}{\bibfnamefont{N.~S.} \bibnamefont{Wingreen}},
  \bibinfo{journal}{Phys. Rev. Lett.} \textbf{\bibinfo{volume}{68}},
  \bibinfo{pages}{2512} (\bibinfo{year}{1992}).

\end{thebibliography}

\end{document}